# Exploration of tetrahedral structures in silicate cathodes using a motif-network scheme

Xin Zhao[1], Shunqing Wu[2], Xiaobao Lv[3], Manh Cuong Nguyen[4], Cai-Zhuang Wang[4], Zijing Lin[3], Zi-Zhong Zhu[2], and Kai-Ming Ho[1, 4, 5]

[1] Department of Physics and Astronomy, Iowa State University, Ames, Iowa 50011, USA.
[2] Department of Physics, Xiamen University, Xiamen 361005, China.
[3] Department of Physics and Collaborative Innovation Center of Suzhou Nano Science and Technology, University of Science and Technology of China, Hefei 230026, China.
[4] Ames Laboratory, US DOE, Ames, Iowa 50011, USA.
[5] International Center for Quantum Design of Functional Materials (ICQD), Hefei National Laboratory for Physical Sciences at the Microscale, University of Science and Technology of China, Hefei 230026, China.

Correspondence and requests for materials should be addressed to X.Z. (email: xzhao@iastate.edu) or S.Q.W. (email: wsq@xmu.edu.cn)

**Using a motif-network search scheme, we studied the tetrahedral structures of the dilithium/disodium transition metal orthosilicates $A_2MSiO_4$ with A = Li or Na and M = Mn, Fe or Co. In addition to finding all previously reported structures, we discovered many other different tetrahedral-network-based crystal structures which are highly degenerate in energy. These structures can be classified into structures with 1D, 2D and 3D M-Si-O frameworks. A clear trend of the structural preference in different systems was revealed and possible indicators that affect the structure stabilities were introduced. For the case of Na systems which have been much less investigated in the literature relative to the Li systems, we predicted their ground state structures and found evidence for the existence of new structural motifs.**

Li$_2$MSiO$_4$ (M = Mn, Fe, Co) have been the subject of intensive studies as promising Li storage materials because of their high potential capacities, low cost, environmental friendliness and excellent safety characteristics [1-19]. Realizing a two electron exchange per formula in orthosilicates leads to higher capacities (e.g. ~ 331 mAh/g for Li$_2$FeSiO$_4$) than the olivine phosphates where there is only one Li atom per formula unit [2, 3]. In the last decade, much effort has been devoted to the study of different Li$_2$MSiO$_4$ polymorphs. However, it was reported that Li$_2$FeSiO$_4$ exhibits a reversible capacity of only 130 ~ 165 mAh/g [1, 4, 5] or high initial charge capacities (~ 240 mAh/g) with noticeable decay in the following cycles [6, 7], while both Li$_2$MnSiO$_4$ [2, 8-10] and Li$_2$CoSiO$_4$ [11] show more than one electron exchange in the first charge cycle but suffer from poor rate capability and drastic capacity fade.

In comparison with the Li compounds, much less experimental work was carried out to investigate the orthosilicates as Na host matrix. The chemical similarities between Na and Li imply that exploration of the sodium equivalent offer more opportunities to advance energy storage technology through rechargeable batteries, owing to the even lower cost and ubiquitous availability of Na. Recently [20], Na$_2$MnSiO$_4$ was synthesized and investigated for use as a positive electrode material for Na secondary batteries. A reversible capacity of 125 mAh/g was found compared with the theoretical capacity of 278 mAh/g based on the two electron reaction.

The discrepancy between measured and calculated capacities has been attributed to the instability of the crystal structures upon delithiation/desodiation [3, 12, 13, 20]. In order to circumvent the capacity fading and further improve the electrochemical properties, it is essential to understand their crystal structures and explore other possible polymorphs that may be stable in the delithiated/desodiated state.

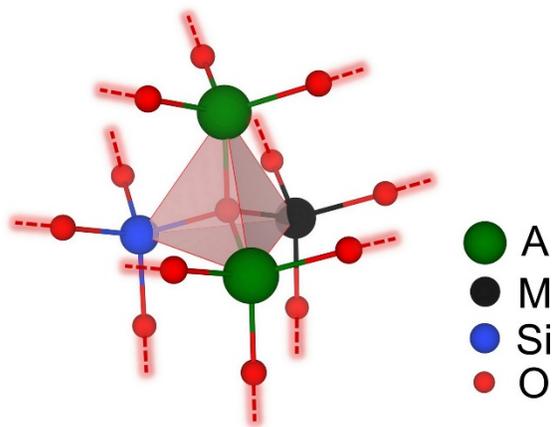

Figure 1 **Schematic representation of the structure generations.** The $A_2MSiO_4$ structures are generated from tetrahedral networks, where A = Li or Na; M = Mn, Fe or Co. For a given tetrahedral network, once one of its sites (e.g. the center of the tetrahedron) is assigned to oxygen, its four neighbors are randomly assigned to two A atoms, one M atom and one Si atom. Then, neighbors of A, M and Si are only assigned to oxygen atoms. In such an iterative manner, the occupations of all sites are determined. The oxygen-centered tetrahedron is shown by red, transparent planes.

Experimental data indicate that the crystal structures of the orthosilicate compounds $A_2MSiO_4$ (A = Li, Na; M = Mn, Fe, Co) belong to a family of tetrahedral structures that exhibit a rich polymorphism [21, 22]. Polymorphs of these tetrahedral structures were classified into low- and high-temperature forms, which differ in the distribution of cations within tetrahedral sites of a hexagonal close-packed (hcp) based arrangement of oxygen. Five different structures were observed and studied for $Li_2FeSiO_4$ [1, 4, 5, 14-16], three as-synthesized (two are orthorhombic, *Pmnb* and *Pmn*$2_1$; one is monoclinic, *P*$2_1$/*n*) and two cycled phases (*Pmn*$2_1$-cycled and *P*$2_1$/*n*-cycled). Likewise, multiple phases have been reported for $Li_2MnSiO_4$ (*Pmn*$2_1$ [2], *Pn* [12], *P*$2_1$/n [17] and *Pmnb* [9]) and $Li_2CoSiO_4$ (*Pnb*$2_1$ [18], *Pmn*$2_1$ [11, 18], and *P*$2_1$/*n* [18]). The recent work of $Na_2MnSiO_4$ [20] showed that $Na_2MnSiO_4$ has a monoclinic structure with space group *Pn*.

In the above reported structures of $A_2MSiO_4$, all the atoms form tetrahedral units, i.e. every atom is in the center of a tetrahedron and has a coordination number of 4. Taking advantage of this structural feature, we used a fast motif-network scheme based on genetic algorithm (GA) [23] to explore the complex crystal structures of these materials. Our results provide a more comprehensive tetrahedral structure database to assist future effort on the study of delithiation/desodiation process.

Although systematic enumerations of 4-connected crystalline networks have been applied to zeolites and other silicates [24-26], considering the great effort of selecting energetically preferable structures out of millions of possible configurations owing to the lack of decent classical potentials for $A_2MSiO_4$, here we took a different route to obtain tetrahedral networks from the low-energy crystal structures of silicon. Silicon is well known to have rich phases and forms sp3-hybridized framework structures [27]. We used GA and Tersoff potential [28] to search for silicon structures that form tetrahedral networks. Once such a silicon structure was located, all the sites were re-assigned to A (Li or Na), M (Mn, Fe or Co), Si and O atoms in the ratio of 2:1:1:4. During the substitution, only structures where every oxygen atom bonds with two A atoms, one M atom and one Si atom, as illustrated in Fig. 1, were accepted. This is because of the observation that structures with uniformly distributed A, M and Si atoms have noticeably lower energies. Newly generated structures that had not been visited were collected for further refinement by first-principles calculations. In this way, various $A_2MSiO_4$ structures were obtained. More details on the first-principles calculations can be found in the methods section.

Generation of the tetrahedral networks costs very little time due to the usage of classical potentials during the GA searches. In this work, up to 48 atoms in the unit cell were searched for Si to find tetrahedral networks, i.e. up to 6 formula units were considered for $A_2MSiO_4$. In order to obtain as many tetrahedral networks as possible, energies of the silicon structures that satisfy the coordination constraints (every atom in the structure has a coordination number of 4) were lowered by a pre-set amount to increase their chance of survival.

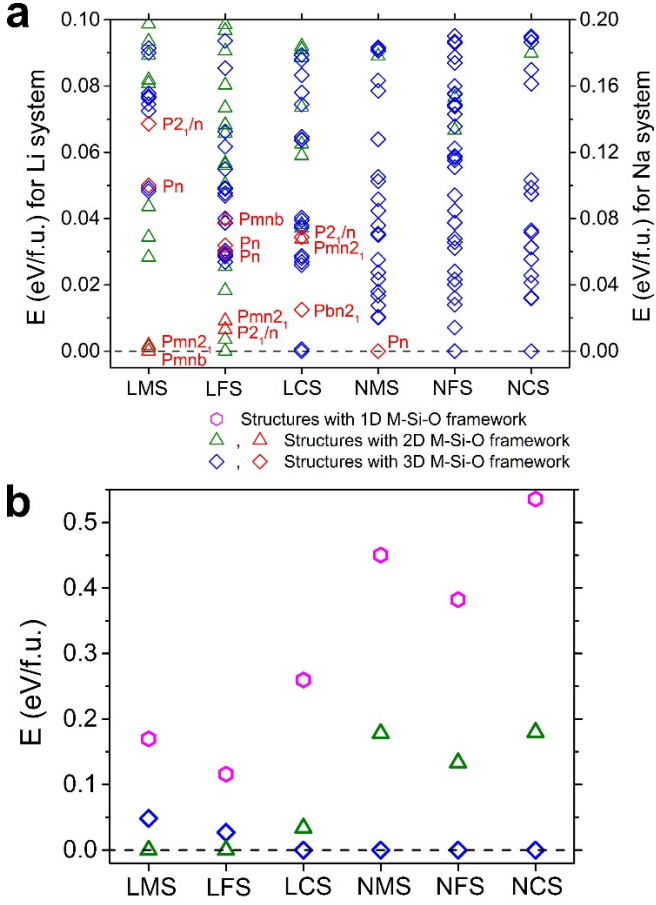

Figure 2 **Energetics results. a,** Relative energies of the structures obtained in this work for Li$_2$MnSiO$_4$ (LMS), Li$_2$FeSiO$_4$ (LFS), Li$_2$CoSiO$_4$ (LCS) and Na$_2$MnSiO$_4$ (NMS), Na$_2$FeSiO$_4$ (NFS), Na$_2$CoSiO$_4$ (NCS). Triangles (green) indicate the structures with layered 2D-framework and diamonds (blue) indicate the structures with 3D M-Si-O framework. Structures that have been reported in the literature are shown in red color and also labeled by their space groups. For the two LFS *Pn* phases, the lower-energy one corresponds to the *Pmn*2$_1$-cycled phase with 2 formula units and the higher-energy one corresponds to the *P*2$_1$/*n*-

cycled phase with 4 formula units. **b**, Relative energies of the most stable structures with 3D, 2D and 1D M-Si-O framework for each system. Energy of the ground state structure for each system is set to 0 eV as reference in **a** and **b**.

Results of the $A_2MSiO_4$ structures from current study are summarized in Fig. 2, where the relative energies are plotted by setting the energy of the ground state structure to 0 eV for each system. We found that the structures of $A_2MSiO_4$ are highly degenerate in energy, in agreement with the rich crystal chemistry observed in experiments. Using our method, in addition to the structures previously reported in the literature (shown in red color in Fig. 2a) and structures included in the Materials Project database [29], many more structures with competitive or even lower energies were found. Within the energy windows plotted in Fig. 2a, less than 10 structures were included in the Materials Project database for each Li system and none for the Na systems, while more than 30 structures are shown in Fig. 2a for each system. We classified those low-energy structures into three different types based on the frameworks formed by M, Si and O atoms [13, 15], i.e. the structures with 1D/2D/3D M-Si-O framework.

**Structures with 3D M-Si-O framework**

In the first type (referred to as "Structure with 3D M-Si-O framework" from now on), M, Si and O atoms form a 3D framework (see examples plotted in Fig. 3). Difference between the structures in Fig. 3a, 3b and 3c comes from the different orientations of the tetrahedra and all three structures consist of only 2-hole ring as indicated by the arrows in Fig. 3a. In contrast, structures in Fig. 3d and 3e consist of a combination of 1-hole ring and 3-hole ring as indicated in the plot. Structure in Fig. 3f mixes the 2-hole rings and the combination of 1 & 3-hole rings. In these structures, M and Si atoms occupy different tetrahedron centers in an hcp sublattice of oxygen, resulting in the different orientations displayed in Fig. 3a-f. We believe more structures with similar features and various mixings can be constructed by increasing the size of the unit cell.

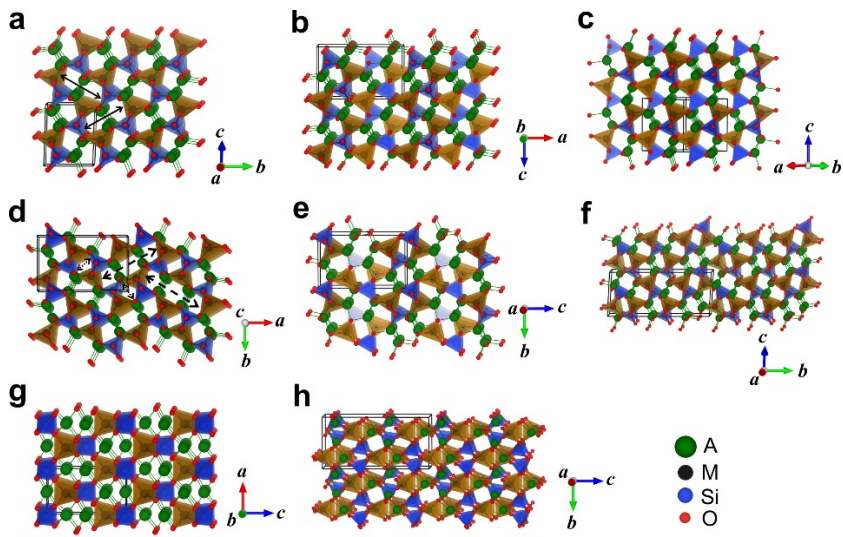

Figure 3 **Examples of the structures with 3D M-Si-O framework.** Space group of each structure is **a,** *Pn* (# 7), **b**, *Pna*$2_1$ (# 33), **c**, *C*$222_1$ (# 20), **d**, *Pna*$2_1$ (# 33), **e**, *P*$2_12_12_1$ (#19), **f**, *Pn* (# 7), **g**, *I*-4 (# 82), and **h**, *Pccn* (# 56). Solid arrows in **a** indicate the 2-hole ring; dash arrows in **d** indicate the 3-hole ring; dot arrows in **d** indicate the 1-hole ring. The black boxes indicate the unit cells of each structure. M- and Si- centered tetrahedra are displayed in the brown and blue colors respectively. A-O bonds are connected and displayed in green color.

The structures plotted in Fig. 3g and 3h look distinct from the others, but the M and Si atoms share the same local tetrahedral environment. Although less favored in energy than the structures plotted in Fig. 3a-3f, the differences are very small. For instance, for Na$_2$FeSiO$_4$, the energies of the structures in Fig. 3g and 3h are about 0.11 and 0.12 eV/f.u. higher than the ground state structure, respectively.

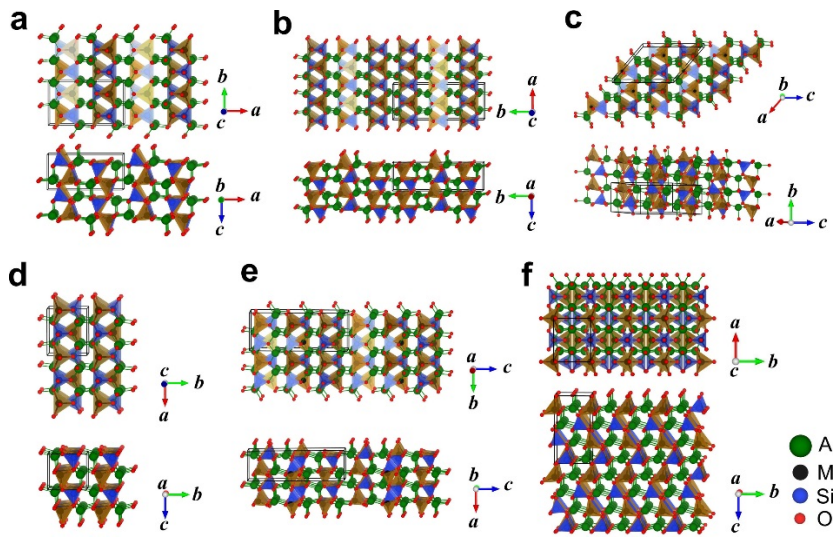

Figure 4 **Examples of the structures with 2D M-Si-O framework.** Space group of each structure is **a**, *Pnma* (# 62), **b,** *Pmn2$_1$* (# 31), **c,** *P2$_1$/n* (# 14), **d,** *Pmn2$_1$* (# 31), **e,** *P2$_1$/m* (# 11), **f**, *Pn* (# 7). Two mutually perpendicular views are plotted for each structure. The black boxes indicate the unit cells of each structure. M- and Si- centered tetrahedra are displayed in the brown and blue colors respectively. A-O bonds are connected and displayed in green color.

**Structures with 2D M-Si-O framework**

The second type (referred to as "Structure with 2D M-Si-O framework" from now on) is that M, Si and O atoms form disconnected layers, as those plotted in Fig. 4. Similar to the structures with 3D M-Si-O framework, M and Si atoms can occupy different tetrahedron centers and as a result, the orientation of the tetrahedra looks different in different structures. For example, the structures plotted in Fig. 4a, 4b and 4d are from various stacking of two different tetrahedron-oriented layers and in each layer, all the tetrahedra point to the same direction. In comparison, layers in the structures plotted in Fig. 4c and 4e mix different-oriented tetrahedra. It can also be expected that by increasing the unit cell size, more ways to stack those layers can be found. Meanwhile, through the exchange of the A and M atoms, more layered structures were found as Fig. 4f, which becomes closer to the structures with 3D M-Si-O framework.

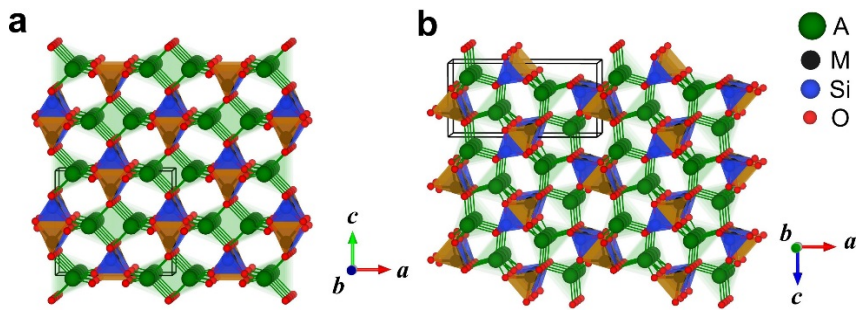

Figure 5 **Examples of the structures with 1D M-Si-O framework.** The structure plotted in **a** has space group *Cmcm* (#63) and the structure plotted in **b** has space group *Pnma* (# 62). The A-, M- and Si-centered tetrahedra are plotted in the color of green, brown and blue respectively. Black boxes indicate the unit cells of each structure.

**Table 1.** Lowest-energy structures of $A_2MSiO_4$ in three different types obtained in current study. r represents the atomic radius; E is the total energy in eV/f.u.; V is the volume of the structure in $Å^3$/f.u.; a, b, and c are the lattice parameters in Å. The corresponding figure of each structure is listed in the "plot" row.

| | | $Li_2MnSiO_4$ | $Li_2FeSiO_4$ | $Li_2CoSiO_4$ | $Na_2MnSiO_4$ | $Na_2FeSiO_4$ | $Na_2CoSiO_4$ |
|---|---|---|---|---|---|---|---|
| r(A)/r(M) | | 1.04 | 1.07 | 1.10 | 1.18 | 1.22 | 1.25 |
| Structures with 1D M-Si-O framework | E | -54.891 | -53.174 | -51.070 | -52.212 | -50.497 | -48.398 |
| | Space group | *Cmcm* (#63) | *Cmcm* (#63) | *Cmcm* (#63) | *Cmcm* (#63) | *Cmcm* (#63) | *Cmcm* (#63) |
| | Lattice | a=7.40, b=7.56, c=6.42 | a=7.47, b=7.49, c=6.30 | a=7.54, b=7.52, c=6.18 | a=8.89, b=8.09, c=6.39 | a=8.95, b=7.96, c=6.31 | a=8.96, b=7.98, c=6.22 |
| | V | 89.80 | 88.12 | 87.61 | 114.89 | 112.38 | 111.18 |
| | plot | Fig. 5a | Fig. 5a | Fig. 5a | Fig. 5a | Fig. 5a | Fig. 5a |
| Structures with 2D M-Si-O framework | E | **-55.061** | **-53.290** | -51.296 | -52.484 | -50.746 | -48.754 |
| | Space group | *Pmna* (#62) | *Pmna* (#62) | *Pmn2₁* (#31) | *P-1* (#2) | *P-1* (#2) | *P-1* (#2) |
| | Lattice | a=10.91, b=6.38, c=5.10 | a=10.80, b=6.33, c=5.05 | a=6.20, b=5.46, c=5.00 | a=5.61, b=6.11, c=6.27, α=77.64°, β=89.96°, γ=89.87° | a=5.73, b=6.05, c=6.12, α=75.43°, β=87.99°, γ=89.17° | a=5.53, b=6.01, c=6.20, α=103.40°, β=90.27°, γ=90.25° |
| | V | 88.66 | 86.33 | 84.60 | 105.00 | 102.76 | 100.36 |
| | plot | Fig. 4a | Fig. 4a | Fig. 4d | Fig. 7 | Fig. 7 | Fig. 7 |
| Structures with 3D M-Si-O framework | E | -55.012 | -53.263 | **-51.330** | **-52.662** | **-50.879** | **-48.934** |
| | Space group | *Pna2₁* (#33) | *P2₁2₁2₁* (#19) | *Pn* (#7) | *Pn* (#7) | *Pn* (#7) | *Pn* (#7) |
| | Lattice | a=11.05, b=6.39, c=5.07 | a=11.02, b=6.29, c=5.07 | a=5.01, b=16.20, c=8.07, β=128.33° | a=5.42, b=5.72, c=8.87, β=127.39° | a=5.41, b=5.71, c=8.74, β=127.67° | a=5.34, b=5.58, c=8.82, β=127.06° |
| | V | 89.48 | 87.72 | 85.59 | 109.33 | 106.71 | 104.78 |
| | plot | Fig. 3d | Fig. 3e | Fig. 3f | Fig. 3a | Fig. 3a | Fig. 3a |

**Existence of the structures with 1D M-Si-O framework?**

Both the structures with 2D and 3D M-Si-O frameworks have been observed in experiments for Li$_2$MSiO$_4$ and extensively studied in the literature [4, 5, 11-19]. It is natural to continue the query of the existence of "the structure with 1D M-Si-O framework", where the M, Si and O atoms form disconnected rods. From our search, such structures were observed as shown in Fig. 5. In both structures plotted in Fig. 5, the M-centered and Si-centered tetrahedra are edge-sharing with each other and extend along one direction to form the M-Si-O rod. However, the orientations of the M-Si-O rod are different between them, which can be seen by comparing Fig. 5a and 5b. From the view of the Na-centered tetrahedra, we see that in the *Cmcm* structure (Fig. 5a), A and O atoms also form separated rods which align perpendicularly to the M-Si-O rods, while in the *Pnma* structure (Fig. 5b), A and O atoms forms 2D layers. In fact, the *Pnma* structure plotted in Fig. 5b can be obtained from the structure plotted in Fig. 4a by switching all the alkali metal atoms with M and Si atoms and arranging M and Si in an orderly manner.

Under above classification, different symbols are used in Fig. 2a to represent the types of those low-energy structures obtained in this work. It can be seen that within the energy windows plotted in Fig. 2a, i.e. 0.1 eV/f.u. for Li systems and 0.2 eV/f.u. for Na systems, more structures with 2D M-Si-O framework are found for the Li systems and more structures with 3D M-Si-O framework are found for the Na systems. The structures with 1D M-Si-O framework are not showing in Fig.

2a due to their relatively higher energies (0.1~0.2 eV/f.u. for Li-systems and 0.2~0.4 eV/f.u. for Na-systems). In Fig. 2b, we plotted the relative energies of the most stable structures with 3D, 2D and 1D M-Si-O framework for each system, from which the stabilities of each type can be compared. The preference of different structure types for different systems will be discussed next.

**Structure preference and analysis**

In table 1, we listed the lowest-energy structures for each system in three different types. We note that the structures with 2D M-Si-O framework are the ground state for $Li_2MnSiO_4$ and $Li_2FeSiO_4$ while the structures with 3D M-Si-O framework are more favored by $Li_2CoSiO_4$. For the Na-system, all three favor the structures with 3D M-Si-O framework. The trend can also be seen clearly from Fig. 2b. This could be related to the atomic size of the cations. By comparing the atomic radius r of A and M atoms [r(Na) > r(Li) > r(Mn) > r(Fe) > r(Co)], we see that with r(A)/r(M) getting closer to 1, layered structures are more favored. When the atomic size difference between A and M is too big, layered structures will introduce large strain, thus becoming less favored.

On the other hand, it can be seen from Fig. 6a that when the A-O bond length is smaller than the M-O bond length, the structures with 2D M-Si-O framework are favored; otherwise, the structures with 3D M-Si-O framework are favored. Thus the relative bond length between A-O and M-O can serve as a clearer indicator. At the same time, we see that Si-O bond length are very close for all six systems and the changes in A-O bond lengths among different transition metal systems are also small for both Li and Na. In the Na systems, the variance (standard deviation) of the bond length from the mean value is significantly larger than the Li system, i.e. larger distortions are found in the Na systems due to the larger size of the Na atom. As a result, in comparison with $Li_2MSiO_4$, the structures of $Na_2MSiO_4$ have relatively lower symmetries.

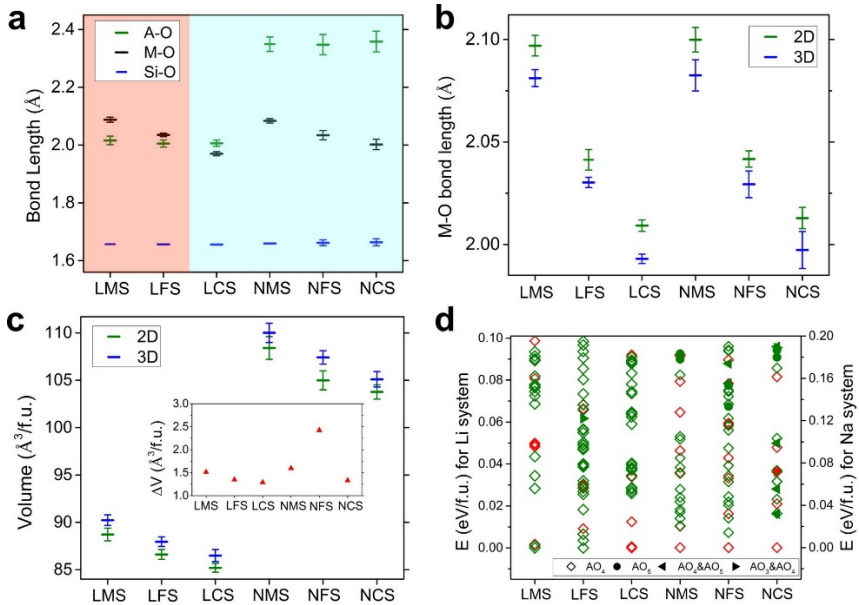

Figure 6 **Structure analyses. a**, Average cation-oxygen bond lengths in different systems. The average is calculated over 30 lowest-energy structures for each system. The red-shaded area represents systems favoring the structures with 2D M-Si-O framework and the blue-shaded area represents systems favoring the structures with 3D M-Si-O framework. **b,** Average M-O bond lengths in the structures with 2D and 3D M-Si-O framework for different systems. **c,** Average volumes of the structures with 2D and 3D M-Si-O framework for different systems. The average volume difference is plotted as the inset. **d,** Local environment of the alkali metal atoms and the connections between the cation-centered tetrahedra in all the structures plotted in Fig. 2a. Green color indicates structures that have edge-sharing tetrahedra; red color indicates structures with only vertex-sharing tetrahedra. Different symbol types represent different local environment of the A (=Li, Na) atoms, i.e. how many oxygen atoms are neighbored by the A atoms. Error bars in plots **a**, **b** and **c** represent one standard deviation of the samples.

To compare the structures with 2D and 3D M-Si-O framework, in Fig. 6b and 6c, we plotted the statistical results of the M-O bond lengths and volumes of them. It is found that for all six systems, the M-O bond lengths in the structures with 2D M-Si-O framework are larger than those in the structures with 3D M-Si-O framework, yet the volumes of the structures with 2D M-Si-O framework are smaller than those of the structures with 3D M-Si-O framework. As for the structures with 1D M-Si-O framework, from the information listed in table 1, it can be seen that the lowest-energy structure with 1D M-Si-O framework for all six systems has space group *Cmcm* with much larger volume than the structures with 2D and 3D M-Si-O framework.

In Fig. 6d, we plotted the local environment of the alkali metal atoms and also the connections between the cation-centered tetrahedra for all the structures in Fig. 2a. To determine whether an oxygen atom is counted as a nearest neighbor of the cation atom, we first sorted all the cation's neighbors according to distance and allowed 10% of increase in the bond length relative to the average of those which have been counted. The results show that for most $Li_2MSiO_4$ structures, the Li atoms bond with 4 oxygen atoms; while for $Na_2MSiO_4$, Na atoms in some structures have different coordination numbers. As shown in Fig. 6d, Na atoms can have coordination numbers of 3 or 5.

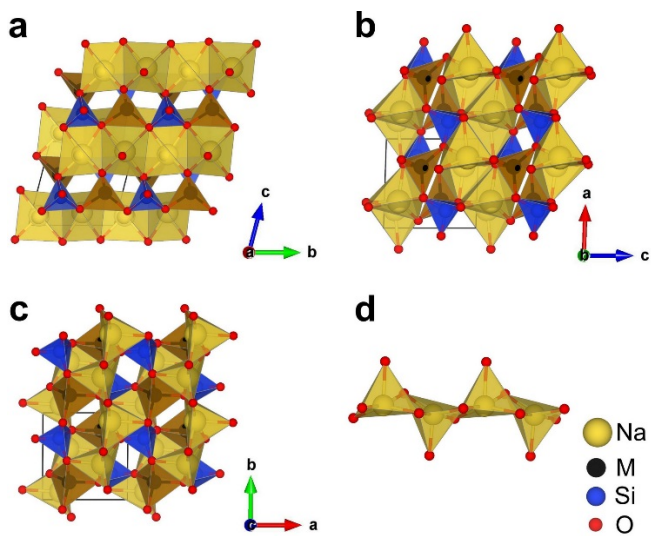

Figure 7 **The lowest-energy structure with 2D M-Si-O framework for the Na systems with space group *P*-1 (#2). a, b, c,** Views of the *P*-1 structure along different lattice vectores. **d,** Na-O pyramids extracted from this structure, where every Na atom bonds with 5 O atoms.

Among all the low-energy structures, we also find that most of them contain edge-sharing tetrahedra which are shown in the green color in Fig. 6d. Structures with only vertex-sharing tetrahedra, as shown in the red color, are more common in the Na systems, but overall, there is no clear indication on how the connection of tetrahedra affects the stability of the structures. (The analysis of deviations from ideal tetrahedral coordination and the corresponding structure files are also given as the supplementary materials.)

**What can be expected for the Na systems?**

Since the Na-intercalation chemistry of the Na-based systems has been considerably less explored, there may be opportunity to find novel electrode materials for sodium-ion battery [30]. Experimental studies on the orthosilicates as Na host matrix have just begun.

In this work, we found that Na systems prefer structures with 3D M-Si-O framework and have relatively low symmetries. As shown in table 1, the lowest-energy structure for all the Na systems has space group *Pn* and similar lattice parameters. The *Pn* structure, which is plotted as Fig. 3a, has been reported for Na$_2$MnSiO$_4$ experimentally [20]. Among the structures with 2D M-Si-O framework obtained in current study, the lowest-energy one for all three Na systems has space group *P*-1. This *P*-1 structure is plotted in detail in Fig. 7. Comparing with those plotted in Fig. 4, the lowest-energy structure with 2D M-Si-O framework for Na system is much more distorted under DFT relaxation and the coordination number of all Na atoms is 5. In Fig. 7d, the Na-O pyramids were plotted. We can see that the center Na atom sits very close to the base plane and four of the five Na neighbors are almost located on the same plane, i.e. such NaO$_5$ pyramid can be considered as half of an octahedron.

The much larger distortions observed in the Na systems indicates that structures with brand new motifs and more competitive energies could exist for the Na compounds, which cannot be fully covered using the method presented in this work. The search space starting from tetrahedral networks has been limited and further studies using more general search schemes should be carried out in order to get a more comprehensive picture of the $Na_2MSiO_4$ structures.

**Conclusion**

In conclusion, by taking advantage of known structural features, we developed a fast motif-network scheme to study the complex crystal structures of the silicate cathode systems for Li-ion/Na-ion batteries. Using the tetrahedral networks generated from silicon, we found that the structures of $A_2MSiO_4$ for both Li and Na systems are highly degenerate in energy. All the structures of $Li_2FeSiO_4$, $Li_2MnSiO_4$, $Li_2CoSiO_4$ and $Na_2MnSiO_4$ that have been reported in the literature were successfully found in our search. Many structures with comparable or even lower energies were revealed, and classified into three different types based on the M-Si-O frameworks.

Through statistical analysis, we showed that structure preference can be related to the relative atomic radius of A and M atoms and the relative bond length of A-O and M-O bonds. Based on these factors, the structures of $A_2MSiO_4$ systems may be controlled through alloying, e.g. doping atoms with different sizes. In addition, existence of brand new motif/structure may be expected in such systems, especially for the Na compounds. The scheme proposed here can be easily extended to other similar systems and serve as a novel approach for extensive exploration of complex crystal structures.

**Methods**

The first-principles calculations on $A_2MSiO_4$ (A = Li, Na; M = Mn, Fe, Co) were carried out using the projector augmented wave (PAW) method [31] within density functional theory (DFT) as implemented in the Vienna ab initio simulation package (VASP) [32, 33]. The exchange and correlation energy is treated within the spin-polarized generalized gradient approximation (GGA) and parameterized by Perdew-Burke-Ernzerhof formula (PBE) [34]. Wave functions are expanded in plane waves up to a kinetic energy cut-off of 500 eV. Brillouin zone integration was performed using the Monkhorst-Pack sampling scheme [35] over k-point mesh resolution of 2π×0.03 Å$^{-1}$. The ionic relaxations stop when the forces on all the atoms are smaller than 0.01 eV·Å$^{-1}$.

Since the energy difference between ferromagnetic (FM) and antiferromagnetic (AFM) is very small and the resulting lattice parameters are almost the same [36, 37], all calculations in present work were spin-polarized with FM configuration. The effects due to the localization of the d electrons of the transition metal ions in the silicates were taken into account with the GGA + U approach of Dudarev et al. [38]. Within the GGA + U approach, the on-site coulomb term U and the exchange term J were grouped together into a single effective interaction parameter Ueff=U-J. In our calculations, U-J values were set to 4 eV for M = Fe, and 5 eV for M = Co, Mn, respectively.

**Acknowledgement**

S.Q.W. and Z.Z.Z. acknowledge the financial support from the National Natural Science Foundation of China under grant Nos. 21233004 and 11004165, the Natural Science Foundation of Fujian Province of China (Grant No. 2015J01030), the Fundamental Research Funds for the Central Universities (Grant No. 20720150034) and the National Basic Research Program of China (973 program, Grant No. 2011CB935903). Z.J.L. and X.L. acknowledge the support by the State Key Development Program for Basic Research of China (Grant No. 2012CB215405) and the National Natural Science Foundation of China (Grant No. 11374272). Work at Ames Laboratory was supported by the US Department of Energy, Basic Energy Sciences, Division of Materials Science and Engineering, under Contract No. DE-AC02-07CH11358, including a grant of computer time at the National Energy Research Scientific Computing Center (NERSC) in Berkeley, CA.